\documentclass[aip,jap,amsfonts,amssymb,amsmath,groupedaddress,twocolumn]{revtex4}

\usepackage{graphicx}
\usepackage{amsmath}
\usepackage{amsfonts}
\usepackage{amssymb}
\usepackage{graphicx}
\usepackage{epstopdf}

\begin{document}

\title{Thermoelectric calculations of ring-shaped bands in two-dimensional Bi$_2$Te$_3$, Bi$_2$Se$_3$ and Sb$_2$Te$_3$: a comparison of simple scattering approximations}

\author{Cameron Rudderham}
\affiliation{Department of Physics and Atmospheric Science, Dalhousie University, Halifax, Nova Scotia, Canada, B3H 4R2}
\author{Jesse Maassen}
\email{jmaassen@dal.ca}
\affiliation{Department of Physics and Atmospheric Science, Dalhousie University, Halifax, Nova Scotia, Canada, B3H 4R2}

\begin{abstract}
Materials with ring-shaped electronic bands are promising thermoelectric candidates, since their unusual dispersion shape is predicted to give large power factors. While previous calculations of these materials have relied on the assumption of a constant mean-free-path or relaxation time, recent first-principles modeling of electron-phonon scattering suggests that the scattering rates may be better approximated by the electron density-of-states (so-called DOS scattering model). In this work, we use density functional theory to investigate single and double quintuple-layer Bi$_2$Te$_3$, Bi$_2$Se$_3$ and Sb$_2$Te$_3$, with a focus on understanding how the three aforementioned scattering approximations impact thermoelectric performance -- emphasis is placed on the DOS scattering model. The single quintuple-layer materials possess two ring-shaped valence band maxima that provide an abrupt increase in conducting channels, which benefits the power factor. Additionally, below the band edge a ring-shaped minimum, located between the two maxima, is found to further enhance the thermoelectric performance but only with the DOS scattering model. This comes from a sharp drop in the DOS, and thus scattering, just below the ring-shaped minimum. An analytic octic dispersion model is introduced and shown to qualitatively capture the observed features. The double quintuple-layer materials display notably worse thermoelectric properties, since their dispersions are significantly modified compared to the single quintuple-layer case. The benefits of ring-shaped bands are sensitive to the alignment of the two ring maxima and to the degree of ring anisotropy. Overall, single quintuple-layer Bi$_2$Te$_3$ and Bi$_2$Se$_3$ are most promising, with the DOS scattering model giving the highest power factors.  
\end{abstract}

\maketitle

\section{Introduction}
\label{sec:intro}
Thermoelectric (TE) materials can convert thermal energy into useful electrical power and thus have the potential to recuperate the large, untapped global energy source that is waste heat \cite{Forman2016}. A major goal is to improve the thermoelectric conversion efficiency, which is characterized by its figure-of-merit \cite{Snyder2008} $ZT =S^2 \sigma T/(\kappa_e + \kappa_l)$, where $S$ is the Seebeck coefficient, $\sigma$ the electrical conductivity, $T$ the temperature, and $\kappa_{e/l}$ the electronic/lattice thermal conductivity. In the quest to achieve higher efficiencies, or $ZT$, there are two broad strategies based on lowering the lattice thermal conductivity and increasing the power factor, $PF=S^2\sigma$. The former has lead to high $ZT$ in many cases, for example using nanostructuring \cite{Hochbaum2008,Poudel2008,Biswas2012} or highly anharmonic materials \cite{Delaire2011,Zhao2014,Li2015,Manley2019} to increase phonon scattering and reduce $\kappa_l$. Approaches for improved $PF$ include (among others) distorted electronic states \cite{Heremans2008,Heremans2012}, band convergence \cite{Pei2011,Liu2012,Tang2015}, low electron-phonon coupling materials \cite{Wang2012,Liu2013,Liu2016,Su2018,Zhou2018}, energy filtering \cite{Vashaee2004,Heremans2005,Bahk2013,Bahk2014,Thesberg2016}, modulation doping \cite{Zebarjadi2011,Zebarjadi2013,Neophytou2016} and unusually shaped electron dispersions.

Focusing on this last strategy, the idea is to identify or design materials with unique electronic band structures that benefit power factor, compared to typical effective mass/parabolic dispersions common in semiconductors. Proposed band structures for high $PF$ include, for example, non-parabolic bands \cite{Chen2013}, ``pudding-mold'' band \cite{Usui2013,Mori2013,Usui2017}, ``camel-back'' dispersion \cite{Wang2014}, semimetals \cite{Markov2018,Markov2019}, topologically-protected states \cite{Ghaemi2010,Pal2015}, and ring-shaped bands \cite{Zahid2010,Maassen2013,Wick2015,Zhou2015,Liang2016,Rudderham2020} -- these dispersions have properties that help circumvent the $\sigma$ versus $S$ trade-off \cite{Snyder2008} that limits power factor. Similar improvements have also been proposed in lower dimensional materials, even those with effective mass bands \cite{Hicks1993a,Hicks1993b,Kim2009}. Focusing specifically on ring-shaped bands (also referred to as ``Mexican hat'' or ``warped'' bands), the key characteristic from this unusual dispersion is a finite number of states at the band edge (in the shape of a ring) that results in an abrupt increase in the distribution-of-modes and the transport distribution; the latter being the central quantity for calculating the TE parameters (discussed more below). This feature allows for simultaneously larger $S$ and $\sigma$, compared to a parabolic band \cite{Rudderham2020}. Ring-shaped bands are often displayed in two-dimensional few-layer materials, such as \cite{Wick2015}: GaSe, GaS, InSe, InS, Bi$_2$Te$_3$, Bi$_2$Se$_3$, bilayer graphene with electric field, and elemental Bi.

Most theoretical studies on ring-shaped bands were carried out using density functional theory (DFT), to obtain detailed and accurate descriptions of the electronic states, but often rely on simple and approximate scattering models \cite{Zahid2010,Wick2015,Zhou2015,Liang2016,Diznab2019,Rudderham2020}; the two most common assume either a constant scattering time or constant mean-free-path (MFP). Rigorous DFT-based electron-phonon calculations have shown that the scattering rates are often better approximated by the electron density-of-states (DOS) \cite{Jiang2017,Witkoske2017,Pshenay2018,Wang2018,Askarpour2019,Graziosi2019}, compared to either a constant scattering time or MFP \cite{Askarpour2019}. A recent study, based on analytical band models, showed that the TE properties can depend sensitively on the details of the scattering approach, and that in particular the DOS scattering model (wherein the scattering rates are assumed to be proportional to the DOS) predicted significantly better performance for ring-shaped bands compared to assuming a constant relaxation time or MFP \cite{Rudderham2020}.

Motivated by these recent findings, this study focuses on revisiting the TE properties of some ring-shaped bands using DFT, and comparing results of the three aforementioned scattering models -- with an emphasis on the DOS scattering model. The materials investigated in this work are two-dimensional Bi$_2$Te$_3$, Bi$_2$Se$_3$, and Sb$_2$Te$_3$, in single and double quintuple-layer (QL) form, which possess warped dispersions originating from spin-orbit coupling. Our results show that certain ring-shaped materials have significantly improved TE characteristics with the DOS scattering model, and thus may be better thermoelectrics than previous reports suggest.

The paper is outlined as follows. Section \ref{sec:theory} introduces the theoretical approach and computational details. The results, and accompanying discussions, are presented in Section \ref{sec:results}. Finally, our findings are summarized in Section \ref{sec:summary}.

\section{Theoretical and Computational Approach}
\label{sec:theory}
\subsection{Transport Formalism}
Within the linear transport regime, the electrical conductivity, Seebeck coefficient and electronic thermal conductivity are defined as \cite{Jeong2010}
\begin{align}
\sigma &= \left( \frac{2q^2}{h} \right) I_0, \label{eq:cond} \\  
S &= -\left( \frac{k_B}{q} \right) \frac{I_1}{I_0}, \label{eq:seebeck} \\ 
\kappa_e &= \frac{2k_B^2T}{h} \left( I_2 - \frac{I_1^2}{I_0} \right), \label{eq:kappae}
\end{align}
with the quantity $I_j$ written as
\begin{align}
I_j = \frac{h}{2}  \int_{-\infty}^{\infty} \Sigma(E) \left( \frac{E-\mu}{k_BT}\right)^j \left[-\frac{\partial f_0}{\partial E}\right]\,dE, \label{eq:fermifunc} 
\end{align}
where $\Sigma(E)$ is the transport distribution, $\mu$ the Fermi level, and $f_0$ the Fermi-Dirac distribution. $\Sigma(E)$ is the central quantity as it contains all material properties and is expressed as \cite{Thonhauser2003,Madsen2006,Jeong2010}
\begin{align}
\Sigma(E) = \frac{1}{\Omega}\sum_{k,s,n} v_x^2(k) \, \tau(k) \, \delta(E-\epsilon(k)),
\label{eq:trans_distr} 
\end{align}
where $\Omega$ is the sample volume, $\epsilon(k)$ is the electronic dispersion ({\it i.e.} band structure), $\tau(k)$ is the scattering time and $v_x = (1/\hbar)(\partial\epsilon/\partial k_x)$ is the group velocity along the direction of transport (here assumed to be the $\hat{x}$ direction). The sum is performed over band index $n$, spin state $s$, and all $k$ states in the Brillouin zone (note that the explicit $s$ and $n$ dependence of the quantities in Eq.~(\ref{eq:trans_distr}) is omitted for clarity).

The TE parameters of a given material depend solely on its transport distribution, and thus is a useful function to analyze. As previously discussed \cite{Rudderham2020}, two of the desired features for $\Sigma(E)$ include a large overall magnitude ({\it i.e.} scaling factor), and a highly asymmetry distribution relative to the Fermi level. The former provides a large electrical conductivity, and the latter results in a large (absolute value) Seebeck coefficient. Using these guidelines, one can easily identify which transport distributions are beneficial for TE transport.

Despite its utility, the transport distribution can be difficult to interpret physically. Turning to the Landauer formalism, we find that $\Sigma(E)$ can be written as the product of two physically-intuitive quantities \cite{Jeong2010,Lundstrom2013,Rudderham2020}
\begin{align}
\Sigma(E) = \frac{2}{h} M(E) \lambda(E). \label{eq:landauer_transport}
\end{align}
The first quantity, $M(E)$, is known as the distribution-of-modes (or DOM), and is defined as \cite{Jeong2010}
\begin{align}
M(E) = \frac{h}{4\,\Omega} \sum_{k,s,n} |v_x(k)|\,\delta(E-\epsilon(k)).
\label{eq:modes}
\end{align}
$M(E)$ can be interpreted physically as the number of ``channels'' available for transport, with each channel (or mode) contributing one quantum of conductance, $2q^2/h$. Eq.~(\ref{eq:modes}) counts the number of modes (an integer) per cross-sectional area, which varies with dimensionality -- the units are m$^{-2}$ in 3D, m$^{-1}$ in 2D and unitless in 1D (all cases in this study are in 2D). The second quantity appearing in Eq.~(\ref{eq:landauer_transport}), $\lambda(E)$, is the mean-free-path for backscattering \cite{Jeong2010}
\begin{align}
\lambda(E) &= 2 \, \frac{\sum_{k,s, n} v_x^2(k)\,\tau(k)\,\delta(E-\epsilon(k))}{\sum_{k,s, n} |v_x(k)|\,\delta(E-\epsilon(k))}. \label{eq:mfp}
\end{align}
$\lambda(E)$ is defined as the average distance along the transport direction that an electron with energy $E$ will travel before scattering changes the sign of its $v_x$ component.

\begin{table*}[tbh]
	\centering
	\begin{tabular}{| c || c | c | c || c | c | c |} 
		\hline
		 & Bi$_2$Te$_3$ (1QL) & Bi$_2$Se$_3$ (1QL) & Sb$_2$Te$_3$ (1QL) & Bi$_2$Te$_3$ (2QL)  & Bi$_2$Se$_3$ (2QL)  & Sb$_2$Te$_3$ (2QL)  \\ [0.5ex] 
		\hline\hline
		DFT band gap (eV) & 0.23 & 0.45 & 0.46 & 0.046 & 0.09 & 0.16 \\ 
		\hline
		GW band gap (eV) & 0.64 & 0.90 & 0.82 & 0.06 & 0.24 & 0.25 \\
		\hline
	\end{tabular}
	\caption{DFT and GW band gaps of single and double QL materials: Bi$_2$Te$_3$, Bi$_2$Se$_3$ and Sb$_2$Te$_3$. The GW band gaps are taken from Refs.~\cite{Forster2015,Forster2016}.}
	\label{tab:band_gaps}
\end{table*}

When the relaxation time $\tau(k)$ is only a function of energy (as considered in this work), {\it i.e.} $\tau(k) = \tau(\epsilon(k)) = \tau(E)$, the mean-free-path for backscattering can be expressed as $\lambda(E) = V_\lambda(E) \tau(E)$ \cite{Rudderham2020}. $V_{\lambda}(E)$ is defined as 
\begin{align}
V_{\lambda}(E) = 2 \,\frac{\sum_{k,s,n}v_x^2(k) \,\delta(E-\epsilon(k))}{\sum_{k,s,n}|v_x(k)|\,\delta(E-\epsilon(k))},
\label{eq:V}
\end{align}
and can be interpreted as an average velocity of carriers, with energy $E$, along the transport direction. In this case, the transport distribution takes on the following form
\begin{align}
\Sigma(E) = \frac{2}{h} M(E) V_\lambda(E) \tau(E),
\label{eq:general_transport}
\end{align}
which has a physically transparent interpretation. $\Sigma(E)$ is determined by the number of channels available for transport, $M(E)$, the average carrier velocity, $V_\lambda(E)$, and the average time between scattering events, $\tau(E)$.

Both $M(E)$ and $V_\lambda(E)$ are calculated directly from a material's band structure. The carrier relaxation time, $\tau(E)$, however depends on the particular scattering physics. Rigorous and accurate scattering calculations of electron-phonon and electron-impurity collision processes, for example based on DFT, are possible but fairly computationally intensive \cite{Giustino2007,Qiu2015,Ponce2016,Zhou2016,Pshenay2018,Zhou2018,Askarpour2019,Graziosi2019}. As a result, simple scattering approximations are often adopted. In this work, we compare the results of three scattering models, based on an assumption of a constant mean-free-path, constant relaxation time, and scattering rates proportional to the DOS.

With a constant MFP, $\lambda_0$, the transport distribution takes the following simple form
\begin{align}
\Sigma_{\text{MFP}}(E) &= \frac{2}{h}M(E) \cdot \lambda_0, \label{eq:td_mfp}
\end{align}
where $\lambda_0$ is an adjustable parameter. We will refer to this scattering approach as the MFP model. A constant MFP is physically expected in the case of a 3D parabolic band with acoustic deformation potential scattering \cite{Lundstrom2000}. With a constant scattering time, $\tau_0$, the transport distribution is written as   
\begin{align}
\Sigma_{\text{TAU}}(E) &= \frac{2}{h} M(E) \cdot V_\lambda(E) \cdot \tau_0, \label{eq:td_tau}
\end{align}
where $\tau_0$ is an adjustable parameter. This approximation will be referred to as the TAU model. A constant relaxation time can be justified physically in the case of a 2D parabolic band with acoustic deformation potential scattering \cite{Lundstrom2000}. In both MFP and TAU models, the approximation is reasonable as long as the MFPs and relaxation times are roughly constant within the energy range ($\sim$10\,$k_BT$) where transport occurs.

Lastly, assuming the scattering rates are proportional to the DOS, the transport distribution is expressed as
\begin{align}
\Sigma_{\text{DOS}}(E) = \frac{2}{h} M(E) \cdot V_\lambda(E) \cdot \frac{K_0}{D(E)}, \label{eq:td_dos}
\end{align}
where $K_0$ is an adjustable parameter. (Note that here we choose to place $K_0$ in the numerator, while in some previous studies it appears in the denominator \cite{Witkoske2017,Wang2018,Rudderham2020}.) This approximation will be referred to as the DOS model. Recent first-principles electron-phonon (el-ph) calculations have shown that the rigorous scattering rates are well described by the DOS scattering approximation \cite{Jiang2017,Witkoske2017,Pshenay2018,Wang2018,Askarpour2019,Graziosi2019}. Physically, electron scattering is expected to scale with the number of available final states, and should follow the DOS when the coupling matrix is roughly constant. This is a decent approximation for non-polar el-ph scattering, as well as for polar el-ph and electron-impurity scattering in highly-doped semiconductors in which screening effects are significant \cite{Lundstrom2000,Askarpour2019}. It is worth mentioning that the two physical examples described above for the MFP and TAU models are, in fact, specific cases of DOS scattering. The concept of scattering rates following the electron DOS is not new and goes back to earlier work \cite{Allen1976,Allen1986,Fischetti1991}, and has been adopted in a number of studies \cite{Zhou2011,McKinney2017,Kumarasinghe2019,Putatunda2019}.

In this work, we compare the transport properties arising from the three scattering models with an emphasis on the DOS scattering approximation, which has not previously been used to analyze these materials. We note, however, that there are limited studies of scattering in ring-shaped two-dimensional materials \cite{Das2019}, thus further investigation is needed to confirm the validity of the DOS model for this material class.

\begin{figure*}	
	\includegraphics[width=15cm]{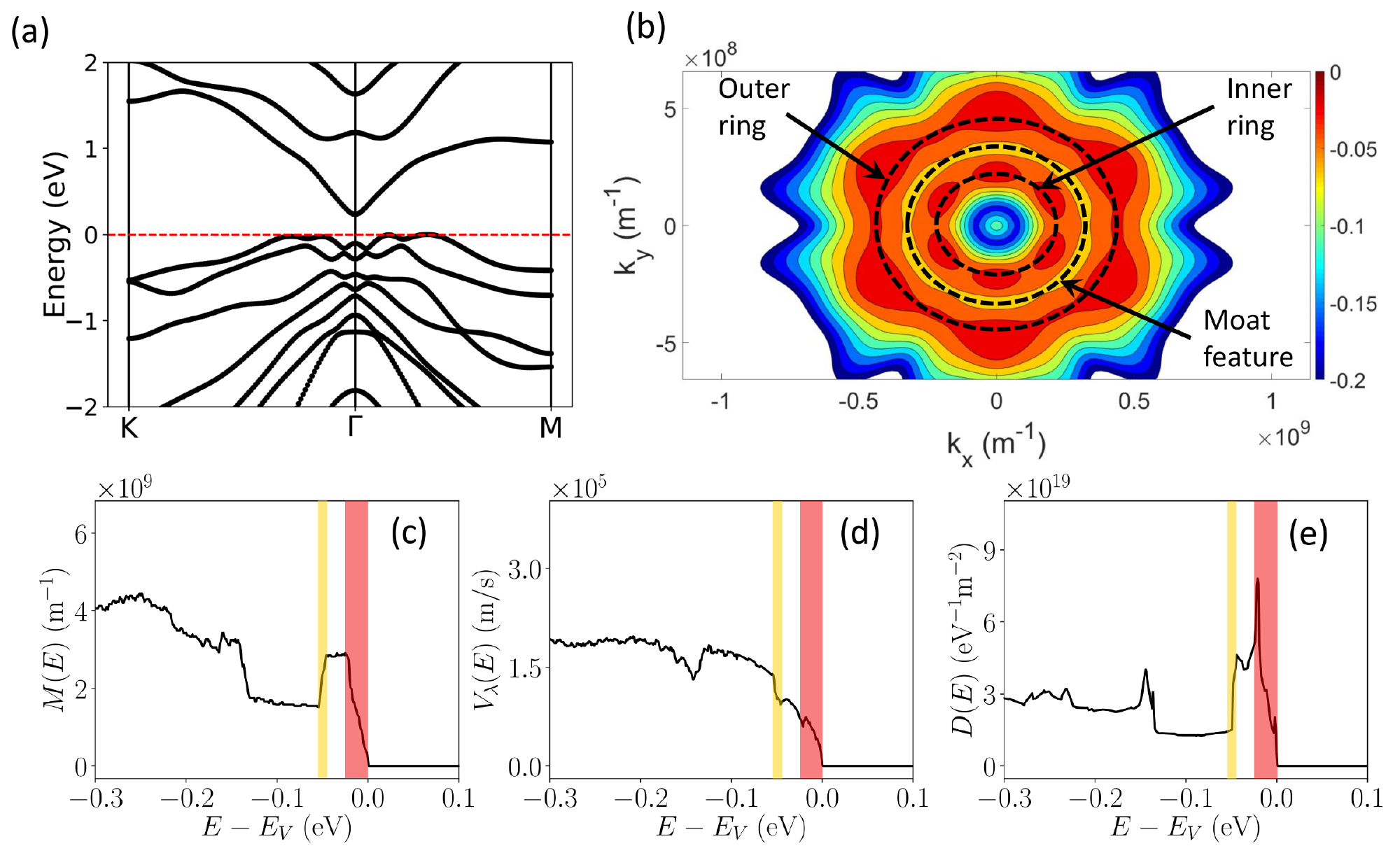}
	\caption{Single quintuple-layer Bi$_2$Te$_3$. (a) Electron dispersion along high-symmetry points. (b) Energy contour plot of valence band. (c) Distribution-of-modes, $M(E)$, (d) average velocity, $V_{\lambda}(E)$, and (e) density-of-states, $D(E)$, versus energy, for the valence states. Shaded regions indicate the energies of the rings and the moat feature.} \label{fig:Bi2Te3_QL}
\end{figure*}

\subsection{Numerical Details}
All DFT calculations were performed with the Quantum Espresso package \cite{QE1,QE2}, using the projector augmented-wave method \cite{paw-method}, the Perdew-Burke-Ernzerhof (PBE) functional of the generalized gradient approximation \cite{Perdew1996}, and fixed occupations. Spin-orbit interaction was included, as well as Grimme-D2 van der Waals corrections \cite{Grimme2006}. A plane-wave cutoff energy of 110~Ry and a Monkhorst-Pack \cite{Monkhorst1976} generated $k$-mesh of 11x11x1 were adopted, for all systems studied. A vacuum layer of 15~$\mathrm{\AA}$, along the $\hat{z}$ direction, was included to prevent interactions between neighboring cells. The experimental lattice constants were used, as has been done previously \cite{Yazyev2010,Liu2010,Maassen2013}, corresponding to an in-plane hexagonal lattice constant of $a$\,=\,4.383~$\mathrm{\AA}$ for Bi$_2$Te$_3$, 4.138~$\mathrm{\AA}$ for Bi$_2$Se$_3$ and 4.264~$\mathrm{\AA}$ for Sb$_2$Te$_3$ \cite{Nakajima1963,Wyckoff1965,Anderson1974}. The atomic coordinates were relaxed until the forces on the atoms were less than 0.01~eV/$\mathrm{\AA}$. In this study, we focus on single and double layer Bi$_2$Te$_3$, Bi$_2$Se$_3$ and Sb$_2$Te$_3$ -- each layer is five-atoms thick (quintuple-layer) and contains strong intraatomic bonds, and the different QLs are held together via weak van der Waals interaction. The primitive cell of these two-dimensional materials is hexagonal with each QL containing five atoms: two equivalent Bi or Sb sites, two equivalent Te or Se sites (at top/bottom surface of each QL), and a third inequivalent Te or Se site (at center of each QL).

The non-self-consistent DOS calculations were performed using the tetrahedron method \cite{Blochl1994} on a uniform 51$\times$51$\times$1 $k$-grid. $M(E)$ and $V_\lambda(E)$ are computed using the ``band-counting'' method \cite{Jeong2010}, which requires the eigenenergies on a uniform $k$-grid with a rectangular Brillouin zone. In this case, the electron energies for a rectangular supercell of size $a$$\times$$\sqrt{3}a$$\times$1 (with double the area of the primitive cell) were calculated with 115$\times$85$\times$1 $k$-points.

\begin{figure*}	
	\includegraphics[width=12cm]{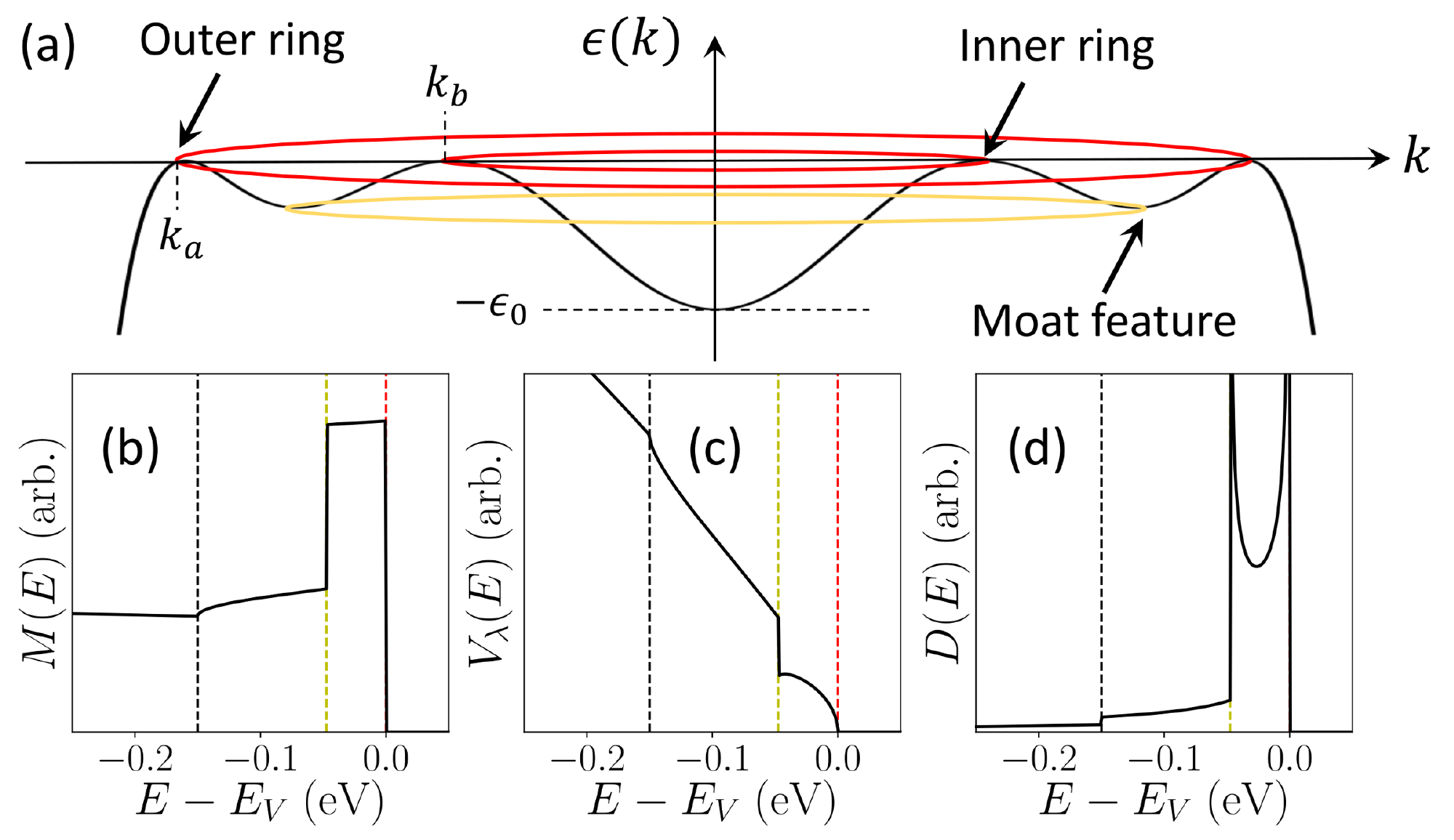}
	\caption{Analytic octic band model. (a) Electron dispersion, $\epsilon(k)$, versus $k$. (b) Distribution-of-modes, $M(E)$, (b) average velocity, $V_{\lambda}(E)$, and (c) density-of-states, $D(E)$, versus energy relative to the valence band edge. To illustrate the shape of the distributions, we used $\epsilon_0$\,=\,0.15~eV, $k_a$\,=\,2 and $k_b$\,=\,1, which gives a moat energy of $-48$~meV.} \label{fig:octic_model}
\end{figure*}

The scissor operator was used to adjust the DFT-calculated band gaps to those obtained from the more accurate GW method \cite{Forster2015,Forster2016} -- see Table~\ref{tab:band_gaps}. All dispersion-related results, including $\epsilon(k)$, $M(E)$, $V_{\lambda}(E)$ and $D(E)$, correspond to the unadjusted DFT band gaps, while the GW band gaps are adopted for the TE transport calculations (to correct for bipolar effects). When evaluating $ZT$, we used a lattice thermal conductivity of 1.5~W/m-K obtained from first-principles phonon transport calculations of single QL Bi$_2$Te$_3$ \cite{Shao2016}. This requires converting our TE parameters from 2D units to 3D units using the following films thickness values: 7.61~$\mathrm{\AA}$ (1QL) and 17.91~$\mathrm{\AA}$ (2QL) for Bi$_2$Te$_3$, 7.04~$\mathrm{\AA}$ (1QL) and 16.60~$\mathrm{\AA}$ (2QL) for Bi$_2$Se$_3$, 7.47~$\mathrm{\AA}$ (1QL) and 17.70~$\mathrm{\AA}$ (2QL) for Sb$_2$Te$_3$. We set the scattering constants (\textit{i.e.} $\lambda_0$, $\tau_0$ and $K_0$) for the valence and conduction states separately, such that the average MFP for backscattering is equal to 20~nm when the Fermi level is located at either the valence ($E_v$) or conduction ($E_c$) band edges (see the Supplemental Information for details on the definition of average MFP for backscattering and how the scattering parameters are determined). All calculations were performed for $T$\,=\,300~K.

\section{Thermoelectric Properties}
\label{sec:results}
We start by analyzing the thermoelectric properties for each of the materials (Bi$_2$Te$_3$, Bi$_2$Se$_3$ and Sb$_2$Te$_3$) in single QL form, then present the results in the case of double QL.

\subsection{Single Quintuple-Layer Bi$_2$Te$_3$}
The band structure for 1QL Bi$_2$Te$_3$ is shown in Fig.~\ref{fig:Bi2Te3_QL}(a). The most significant feature is the presence of not one, but two ring-like features at the valence band edge. From the band structure, this simply looks like two valence band maxima, along both $\Gamma$$\rightarrow$\,M and $\Gamma$$\rightarrow$\,K. However, from the contour plot of the valence band presented in Fig.~\ref{fig:Bi2Te3_QL}(b), we see that the band edges do not correspond to points, but rather have a ring-like shape; specifically a smaller radius (inner) ring and a larger radius (outer) ring. Since this unusual feature in the band structure is the main focus of this study, combined with the fact that our results show that the TE performance of the ``warped'' valence states always surpasses those of the more ``regular'' conduction states, our analysis will focus on the transport properties of the valence states.

The resulting electronic properties, including $M(E)$, $V_\lambda(E)$ and $D(E)$, are shown Fig.~\ref{fig:Bi2Te3_QL}(c)-(e). The ring-like features result in abrupt increases in DOM and DOS near the band edge. While typical electron dispersions, such as the parabolic or Kane models ({\it e.g.} the conduction band of 1QL Bi$_2$Te$_3$), possess a vanishing number of states near the band edge, the valence band of 1QL Bi$_2$Te$_3$ has a large finite number of states within a few meV of the band gap. This gives a sharp step-like feature in the DOM and a spike in the DOS. The large number of states provided by these ring-like features, originating from spin-orbit coupling, is the main reason why these QL materials are promising as high power factor ($PF=S^2\sigma$) thermoelectrics \cite{Zahid2010,Maassen2013,Wick2015,Zhou2015}. $V_\lambda(E)$, by comparison, increases smoothly as we move away from the band edge. This happens because $V_\lambda(E)$ corresponds to an \textit{average} velocity over a constant energy surface (as opposed to a sum like in the case of $M(E)$ or $D(E)$), and because the states near the band edge have small velocities.

\begin{figure}	
	\includegraphics[width=7.0cm]{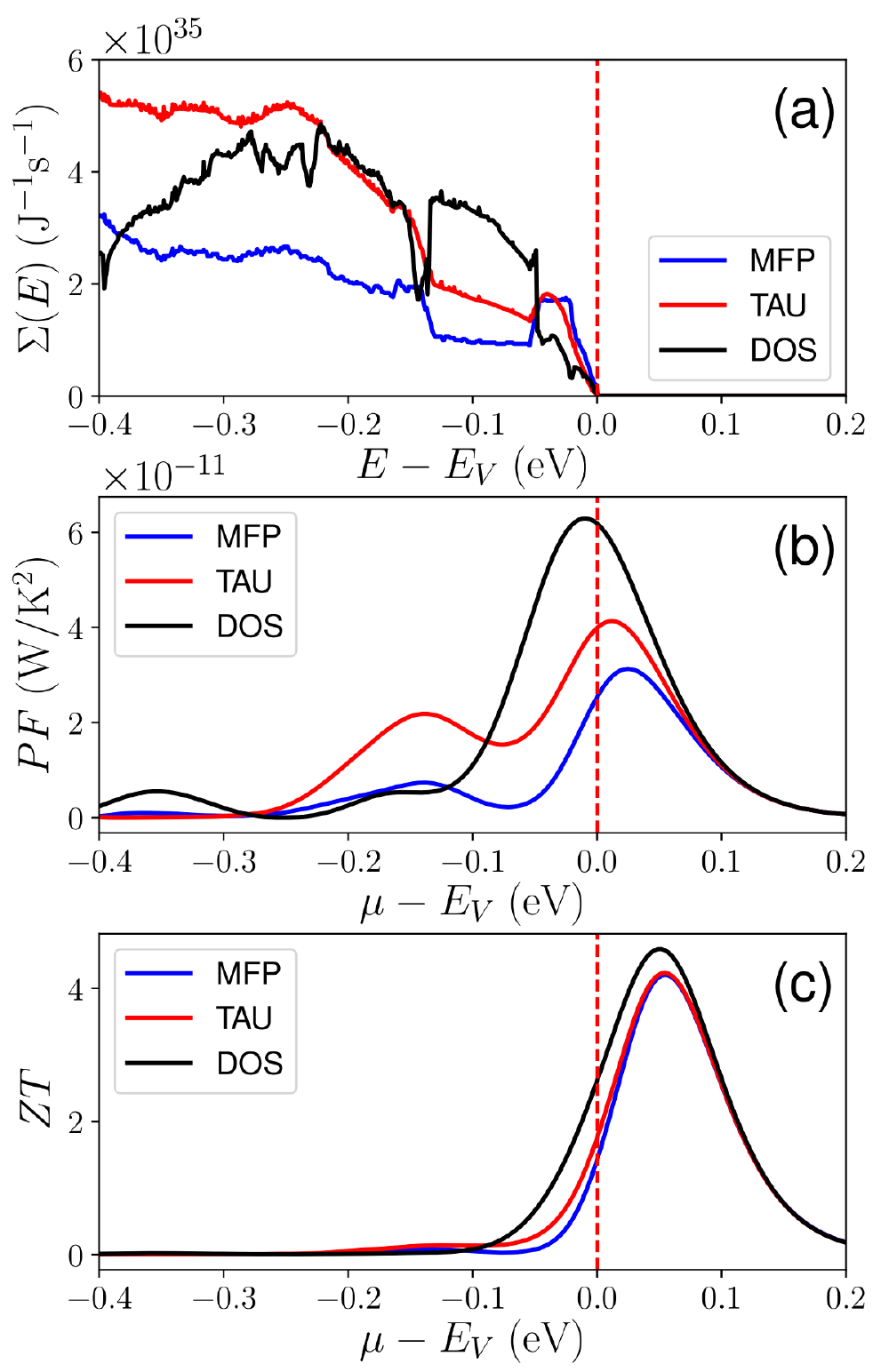} 
	\caption{Single quintuple-layer Bi$_2$Te$_3$. (a) Transport distribution, $\Sigma(E)$, of the valence states versus energy. (b) Power factor, $PF$, and (c) figure-of-merit, $ZT$, versus Fermi level relative to the valence band edge. Comparison of MFP, TAU and DOS scattering models. $T$\,=\,300~K.}
	\label{fig:Bi2Te3_QL_props}
\end{figure}

\begin{figure*}	
	\includegraphics[width=15cm]{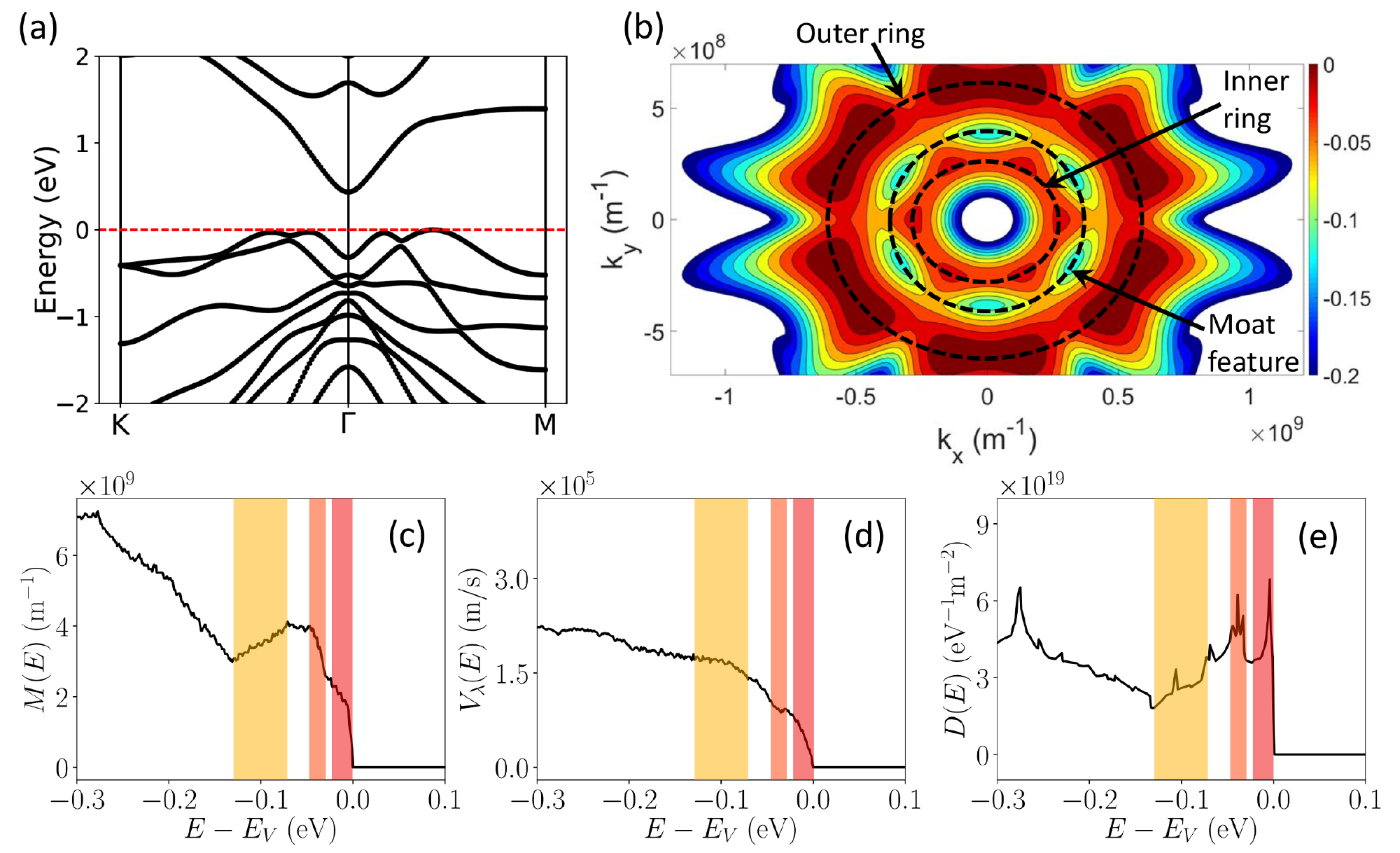}
	\caption{Single quintuple-layer Bi$_2$Se$_3$. (a) Electron dispersion along high-symmetry points. (b) Energy contour plot of valence band. (c) Distribution-of-modes, $M(E)$, (d) average velocity, $V_{\lambda}(E)$, and (e) density-of-states, $D(E)$, versus energy, for the valence states. Shaded regions indicate the energies of the rings and the moat feature.} \label{fig:Bi2Se3_QL}
\end{figure*}

Previously, a 2D analytic dispersion model containing a single ring-like band, known as the quartic or Mexican-hat model, was used to analyze and explore the impact on the TE properties (see Refs.~\cite{Wick2015,Sevincli2017,Rudderham2020} for details). The features in $M(E)$, $V_{\lambda}(E)$ and $D(E)$, discussed above, are all in rough agreement with the quartic model near the band edge. However, slightly away from the band edge, we observe abrupt \textit{decreases} in $M(E)$ and $D(E)$ near 0.05~eV below $E_v$, and a smaller abrupt \textit{increase} in $V_\lambda(E)$. While the quartic model does predict a discontinuous decrease in $D(E)$ at an energy below the band edge, it predicts that $M(E)$ and $V_\lambda(E)$ are continous, and so cannot explain the observed characteristics. To gain further insight into the origin of these features, we examine the contour plot in Fig.~\ref{fig:Bi2Te3_QL}(b).

The presence of the two aforementioned ring-like local maxima at the band edge necessarily requires the existence of a single ring-like local \textit{minimum} nestled between them; a feature observed in Fig.~\ref{fig:Bi2Te3_QL}(b), which we will refer to as the ``moat feature'' (in light of its topographic resemblance to a moat). The bottom of the moat lies at the same energy as the observed discontinuities in $M(E)$, $V_\lambda(E)$ and $D(E)$, suggesting they originate from the disappearance of states below $-0.05$~eV. The near-circular nature of the moat indicates that this particular constant energy line is nearly isotropic.

While the aforementioned quartic band model can describe the impact of ring-like local maxima at the band edge, it does not capture the effect of a ring-like local minima at energies below the band edge. To investigate what effect such a moat feature would have, we introduce a new analytic dispersion model that we refer to as the ``octic model'' (it is an eighth-order polynomial in $k$):
\begin{align}
\epsilon(k) = -\frac{\epsilon_0}{(k_a k_b)^4} \left(k^2 - k_a^2\right)^2 \left(k^2-k_b^2\right)^2. \label{eq:octic}
\end{align}
The octic model possesses three ring-shaped critical lines (two maxima and one local minimum) and a single critical point at $k$\,=\,0, which are illustrated in Fig.~\ref{fig:octic_model}(a). The parameters $k_a$ and $k_b$ correspond to the radii of the critical lines at the band edge ($\epsilon(k)$\,=\,0), {\it i.e.} the inner and outer ring radii. The constant $\epsilon_0$ determines the energy at the $\Gamma$ point: $\epsilon$($k$\,=\,0)\,=\,$-\epsilon_0$. From the three parameters $\epsilon_0$, $k_a$ and $k_b$, the bottom of the moat has an energy of $-\epsilon_0 (k_a^2 - k_b^2)^4/(16 k_a^4 k_b^4)$. For comparison, the quartic model captures a single ring-shaped maximum and a minimum at $k$\,=\,0.

The resulting $M(E)$, $V_\lambda(E)$ and $D(E)$ distributions for the octic model are plotted in Fig.~\ref{fig:octic_model}(b)-(d), with the analytic expressions provided in the Supplemental Information. The presence of ring-shaped critical surfaces at the band edge results in a $M(E)$ distribution that turns on like a step function, just like in the case of the quartic model. This discontinuity is of course sharper than the corresponding feature in the case of 1QL Bi$_2$Te$_3$, since the DFT dispersion shows some small amount of anisotropy resulting in the energy of a ring being spread over a couple meV. In our discussion we will refer to abrupt features, arising from both the analytic model and the DFT-computed results as being ``discontinuous'', even though only the former is technically discontinuous.

The octic model also contains a second discontinuity in $M(E)$ at the location of the moat, in the form of a step-like \textit{decrease}. This occurs because a large number of transport channels, provided by the moat feature, abruptly vanishes. Again, this discontinuity is also observed in 1QL Bi$_2$Te$_3$. Since these states have small velocities near the bottom of the moat, $V_\lambda(E)$ discontinuously increases below the moat, since the average no longer includes a large number of zero-velocity states. A singularity in the DOS occurs at the moat energy, for the same reason as the observed singularity at the band edge; any constant energy containing a continuum of critical points will cause the DOS to diverge (for 2D materials). Overall, the features in $M(E)$, $V_\lambda(E)$ and $D(E)$ obtained from DFT are found to have a strong resemblance with those from the octic model -- confirming that the two ring-shaped maxima in addition to the ring-shaped minimum are key to understanding the observed transport properties.

\begin{figure}	
	\includegraphics[width=7.0cm]{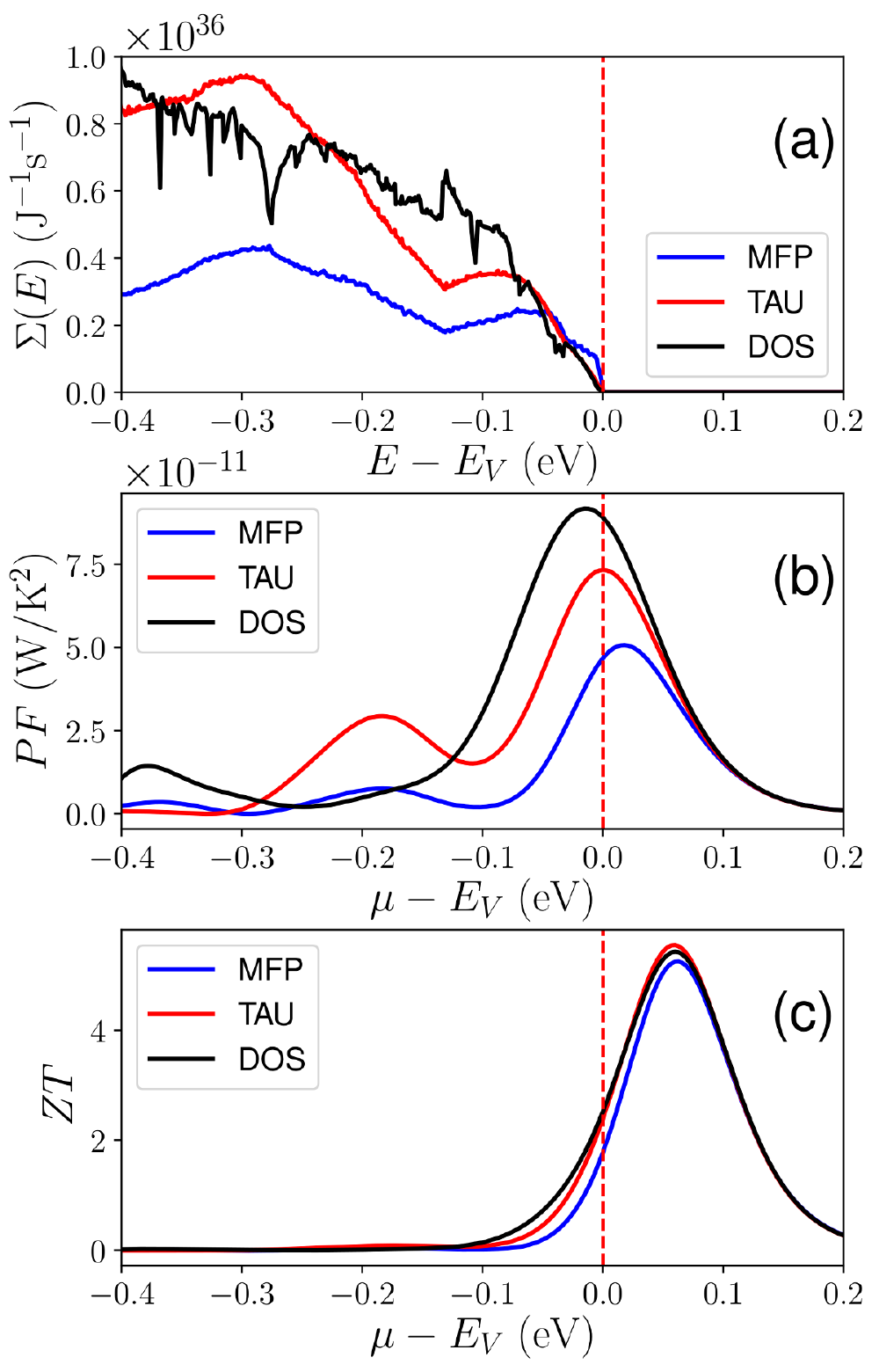}
	\caption{Single quintuple-layer Bi$_2$Se$_3$. (a) Transport distribution, $\Sigma(E)$, of the valence states versus energy. (b) Power factor, $PF$, and (c) figure-of-merit, $ZT$, versus Fermi level relative to the valence band edge. Comparison of MFP, TAU and DOS scattering models. $T$\,=\,300~K.} \label{fig:Bi2Se3_QL_props}
\end{figure}

Next, we examine the transport distribution of 1QL Bi$_2$Te$_3$ for each scattering model (MFP, TAU, DOS), which are shown in Fig.~\ref{fig:Bi2Te3_QL_props}(a). Firstly, we note that $\Sigma(E)$ for each scattering model is qualitatively different at the energy of the moat ($-0.05$~eV). In the case of a constant MFP, the transport distribution is simply proportional to $M(E)$, and hence decreases abruptly just below the moat energy. With the TAU model, the transport distribution is proportional to the product of $M(E)$ and $V_\lambda(E)$; the latter of which steps abruptly upwards once the moat states disappear. The resulting transport distribution still steps downwards at the moat energy, but less than in the constant MFP case.

Lastly, with the DOS model, the transport distribution steps abruptly \textit{upwards} below the moat due to the discontinuity in the $D(E)$ distribution. This results in the DOS model having the largest magnitude in transport distribution, compared to the MFP and TAU models, over most of the relevant energy range (several $k_B T$ around the Fermi level), coupled with a sharp rise with increasing hole energy. Consequently, among the different scattering approximations, the DOS model predicts the largest power factor and $ZT$ values for 1QL Bi$_2$Te$_3$ -- the maximum $PF$ is roughly 100\% and 50\% larger than with the MFP and TAU models, respectively. This improved $PF$ is mainly due to a higher Seebeck coefficient from the sharp rise in $\Sigma_{\rm DOS}(E)$ ($\sigma$ and $S$ for the 1QL materials are presented in the Supplemental Information). This shows that the presence of a moat-like feature can produce an abrupt decrease in $D(E)$ and increase in $V_\lambda(E)$, both of which are desirable for thermoelectric performance. As such, single QL Bi$_2$Te$_3$ may be a better thermoelectric than previously predicted using MFP and TAU models.

A similar result was previously shown with the quartic model, in which an abrupt decrease in DOS (and hence scattering) provided improved power factor when comparing the DOS model to the MFP and TAU models \cite{Rudderham2020}. One difference between the quartic and octic band models is that the decrease in DOS originates from the removal of parabolic-like states and ring-like states, respectively. As a result, only the octic model presents discontinuities in $M(E)$ and $V_\lambda(E)$ below the band edge; these sharper features may be more resilient to small deviations from isotropicity that spread the critical points of the ring over a small energy range (as seen in Fig.~\ref{fig:Bi2Te3_QL}(b)).

\begin{figure*}	
	\includegraphics[width=15cm]{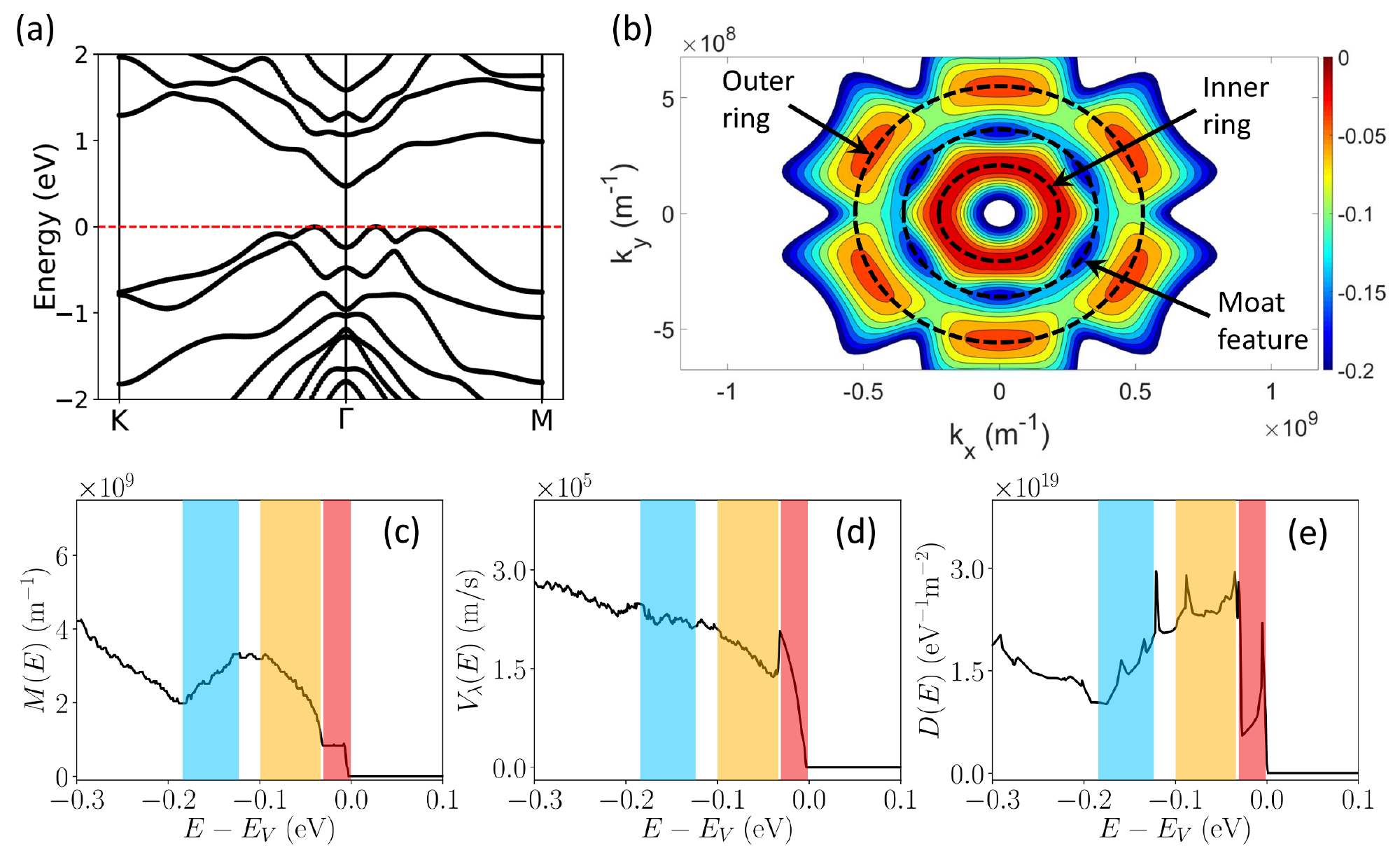}
	\caption{Single quintuple-layer Sb$_2$Te$_3$. (a) Electron dispersion along high-symmetry points. (b) Energy contour plot of valence band. (c) Distribution-of-modes, $M(E)$, (d) average velocity, $V_{\lambda}(E)$, and (e) density-of-states, $D(E)$, versus energy, for the valence states. Shaded regions indicate the energies of the rings and the moat feature.} \label{fig:Sb2Te3_QL}
\end{figure*}

\subsection{Single Quintuple-Layer Bi$_2$Se$_3$}
Next, we analyze the thermoelectric performance of single QL Bi$_2$Se$_3$. The electronic structure of this material is shown in Fig.~\ref{fig:Bi2Se3_QL}(a). We observe that the valence band possesses two ring-like features at the band edge, analogous to 1QL Bi$_2$Te$_3$. As before, this results in a rapid increase in $M(E)$ near the band edge. For this case, however, there are two noticeable ``steps'' in the DOM. This occurs because both rings are slightly misaligned in energy -- the inner ring is roughly 0.03~eV below the outer ring, as seen in Fig.~\ref{fig:Bi2Se3_QL}(b). This also gives a peak in the $D(E)$ at the inner and outer ring energies.

Unlike with 1QL Bi$_2$Te$_3$, however, there are no abrupt features in $M(E)$ and $V_\lambda(E)$ due to the presence of a moat feature. To help understand why, we examine the energy contour plot in Fig.~\ref{fig:Bi2Se3_QL}(b). We observe a moat feature that is highly anisotropic, varying in energy between roughly $-0.12$~eV and $-0.07$~eV, as indicated by the presence of multiple shallow-energy valleys. This anisotropy results in a ``smearing out'' of the sharp features that would have resulted from a more isotropic moat feature, as the abrupt disappearance of a large constant energy surface now happens gradually over an energy range of non-negligible width.

The resulting transport distributions and thermoelectric properties are shown in Fig.~\ref{fig:Bi2Se3_QL_props}. There is a distinct lack of sharp features in any of the transport distributions, compared to 1QL Bi$_2$Te$_3$, as a consequence of the aforementioned anisotropy in the moat feature, in addition to the misalignment in energy of the inner and outer rings. Nevertheless, the DOS-model transport distribution still takes on the largest value over a significant portion of the relevant energy range, and yields the highest $PF$ and $ZT$ among the scattering models. This suggests that 1QL Bi$_2$Se$_3$ may also be a better thermoelectric than previously reported. We note, however, that the relative improvement is not as significant as with 1QL Bi$_2$Te$_3$, indicating that the performance enhancements are sensitive to the degree of anisotropy of the rings/moat and their relative alignment in energy.

The $\Sigma_{\rm DOS}(E)$ for 1QL Bi$_2$Te$_3$ benefits from abrupt increases, which help increase its Seebeck coefficient, however 1QL Bi$_2$Se$_3$ shows larger $PF$ and $ZT$. This happens because the transport distribution of 1QL Bi$_2$Se$_3$, while relatively smooth, has an overall larger magnitude resulting in higher conductivity.

\begin{figure}	
	\includegraphics[width=7.0cm]{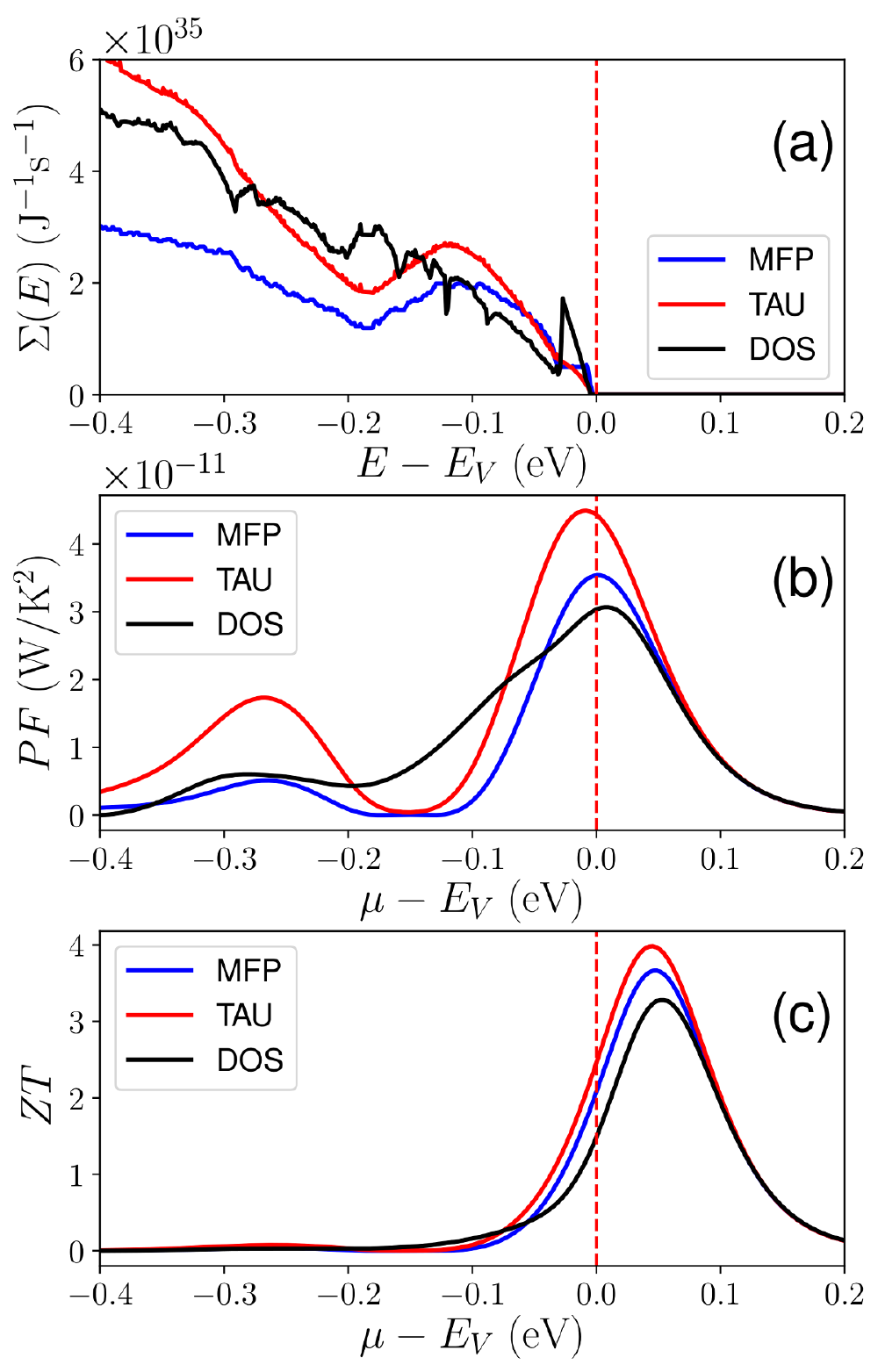}
	\caption{Single quintuple-layer Sb$_2$Te$_3$. (a) Transport distribution, $\Sigma(E)$, of the valence states versus energy. (b) Power factor, $PF$, and (c) figure-of-merit, $ZT$, versus Fermi level relative to the valence band edge. Comparison of MFP, TAU and DOS scattering models. $T$\,=\,300~K.} \label{fig:Sb2Te3_QL_props}
\end{figure}

\subsection{Single Quintuple-Layer Sb$_2$Te$_3$}
The third, and final, single QL material investigated is Sb$_2$Te$_3$. Its electronic structure properties are presented in Fig.~\ref{fig:Sb2Te3_QL}. Similar to the previous two 1QL systems, there are two ring-like features near the valence band edge, however the $M(E)$, $V_\lambda(E)$ and $D(E)$ distributions display sharp features not shared with the other cases. Namely, we observe large increases in the DOM and DOS roughly 0.03~eV below the band edge, accompanied with a sharp decrease in $V_\lambda(E)$. To understand the origin of these features, we examine the energy contour plot presented in Fig.~\ref{fig:Sb2Te3_QL}(b).

As with 1QL Bi$_2$Se$_3$, the inner and outer rings of 1QL Sb$_2$Te$_3$ are slightly misaligned in energy by roughly 0.03~eV. However, the resulting effect on the electronic distributions is somewhat stronger in this case, because the inner ring ``turns on'' before the outer ring ({\it i.e.} the outer ring has lower energy). The outer ring tends to have a stronger effect since its larger radius includes more states compared to the inner ring. The sudden contribution from the outer ring, at $-0.03$~eV, causes the abrupt increases in $M(E)$ and $D(E)$, which are somewhat washed out due to the anisotropy of the outer ring. Because the states near the top of the outer ring have very small velocities, $V_\lambda(E)$ is dragged down. All these features are similar to 1QL Bi$_2$Se$_3$, but are more pronounced in this case.

The effect of this lower-energy outer ring is exactly opposite to that of the moat feature in 1QL Bi$_2$Te$_3$. In the latter case, the abrupt turn-\textit{off} of a ring-like dispersion feature caused sharp decreases in $M(E)$ and $D(E)$, and a sharp increase in $V_\lambda(E)$, whereas in this case the abrupt turn-\textit{on} of a ring-like feature has the opposite effect. Since the moat feature in 1QL Bi$_2$Te$_3$ was shown to benefit TE performance (within the DOS scattering model), we anticipate the misaligned outer ring in 1QL Sb$_2$Te$_3$ may have a detrimental effect. The resulting transport distributions and their corresponding thermoelectric properties are shown in Fig.~\ref{fig:Sb2Te3_QL_props}.

While $\Sigma_{\rm DOS}(E)$ takes on the largest values near the band edge, the increased scattering and decreased velocity that arise from the outer ring at $-0.03$~eV causes the transport distribution to abruptly step down. As a result, it takes on the smallest value of the three transport distributions over a significant portion of the relevant energy range. Moreover, having a segment of the transport distribution decrease, with increasing hole energy, negatively impacts the Seebeck coefficient. Examining the $PF$ and $ZT$ distributions, we observe that the DOS-scattering case predicts the worst performance, for the reasons outlined above.

From our findings on the single quintuple-layer materials, we conclude that the ring-shaped bands, in addition to the moat feature, can benefit TE performance. This mainly comes from abrupt increases in $M(E)$ from the rings at the band edge, coupled with an abrupt decrease in $D(E)$ from the moat feature (which decreases the scattering rates). We also found that these benefits can be sensitive to the relative alignment of the inner and outer rings, and to the degree of anisotropy of the rings/moat that smoothes out the desired sharp features.

\begin{figure*}	
	\includegraphics[width=18cm]{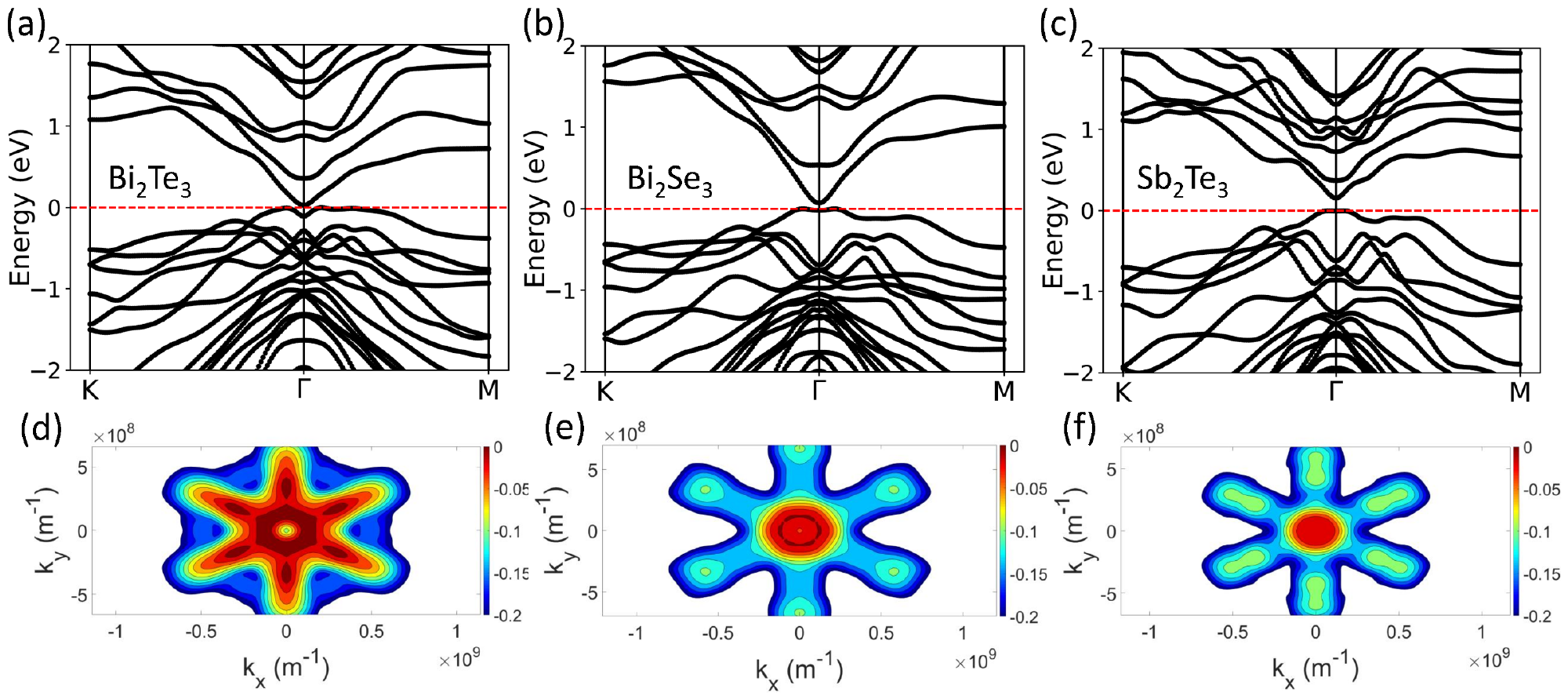}
	\caption{Double quintuple-layer Bi$_2$Te$_3$, Bi$_2$Se$_3$ and Sb$_2$Te$_3$. (a)-(c) Electron dispersion along high-symmetry points. (d)-(f) Energy contour plot of valence band.} \label{fig:2QL}
\end{figure*}

\subsection{Double Quintuple-Layer Bi$_2$Te$_3$, Bi$_2$Se$_3$ and Sb$_2$Te$_3$}
Next, we analyze the thermoelectric properties of Bi$_2$Te$_3$, Bi$_2$Se$_3$ and Sb$_2$Te$_3$ in double quintuple-layer form. The 2QL materials are comprised of two stacked 1QL layers held together via van der Waals interaction. While the interlayer coupling is relatively weak, as we will show, this has a significant effect on the electronic and thermoelectric characteristics.

Figure~\ref{fig:2QL} presents the electron dispersions and energy contour plots for all three 2QL materials. First, we note that all three band gaps are significantly smaller than in the 1QL case. This typically has a negative effect on thermoelectric performance, as it increases bipolar effects. Perhaps more importantly, we observe a dramatic change in the  band structure near the valence band edge. While the 1QL materials display two ring-shaped maxima, in the 2QL systems one of the rings is pushed away from the band edge and loses most of its ring-like character. Both 2QL Bi$_2$Se$_3$ and Sb$_2$Te$_3$ show a small radius ring at the band edge, with lower-energy states near $-0.1$~eV showing a ``starfish''-like shape with six ``arms'' stretching out from the zone center. However, the trend is reversed with 2QL Bi$_2$Te$_3$, which presents a star-shaped dispersion near the band edge with a large radius ring near $-0.15$~eV.

\begin{figure*}	
	\includegraphics[width=15cm]{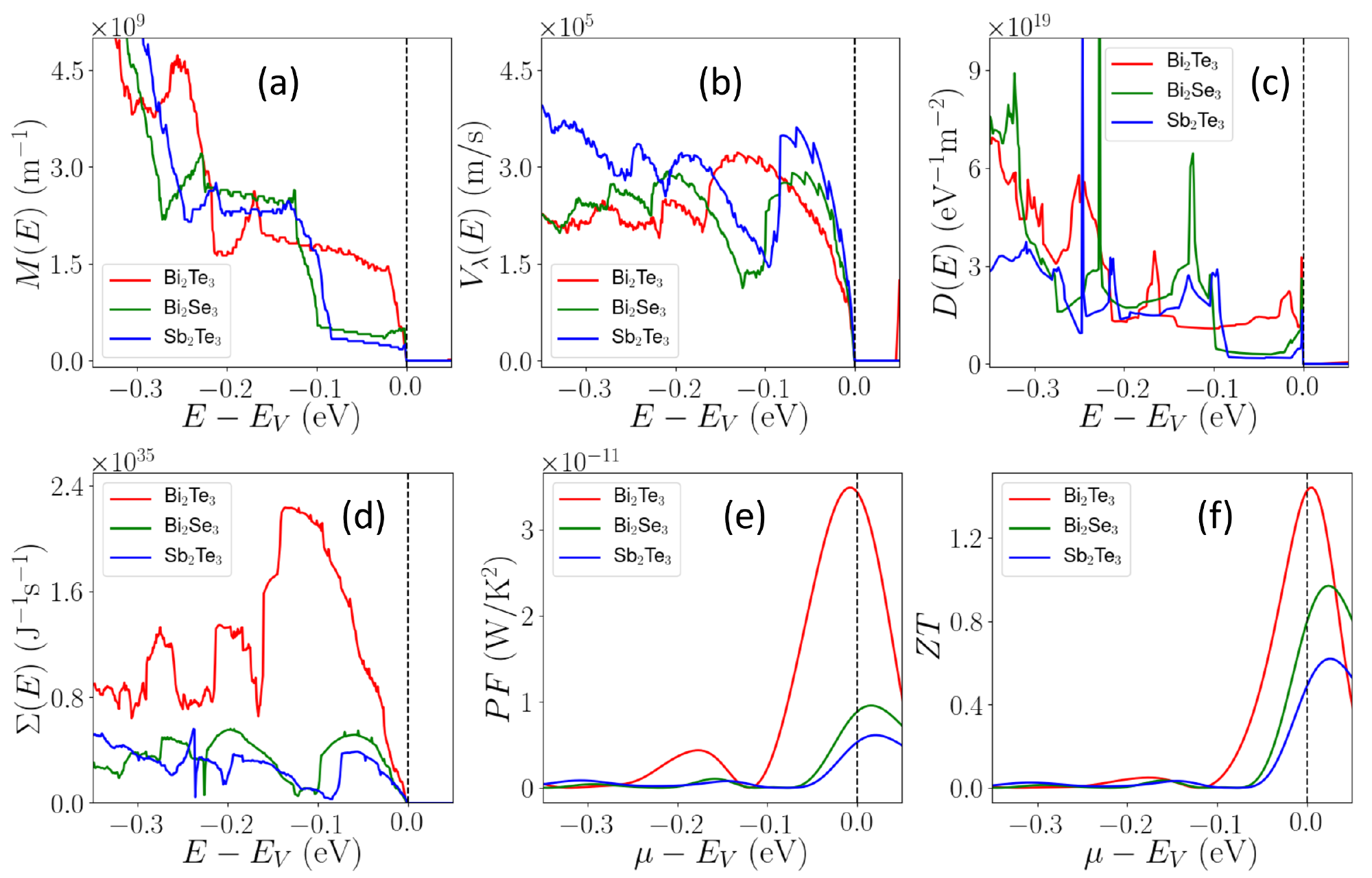}
	\caption{Double quintuple-layer Bi$_2$Te$_3$, Bi$_2$Se$_3$ and Sb$_2$Te$_3$. (a) Distribution-of-modes, $M(E)$, (b) average velocity, $V_{\lambda}(E)$, and (c) density-of-states, $D(E)$, versus energy, for the valence states. (d) Transport distribution, $\Sigma(E)$, versus energy. (e) Power factor, $PF$, and (c) figure-of-merit, $ZT$, versus Fermi level. $\Sigma(E)$, $PF$ and $ZT$ are calculated using the DOS scattering model. $T$\,=\,300~K.} \label{fig:2QL_props}
\end{figure*}

The distribution-of-modes, $M(E)$, average velocity, $V_\lambda(E)$, and density-of-states, $D(E)$, for the 2QL materials are shown in Fig.~\ref{fig:2QL_props}(a)-(c). Similar to the 1QL case, all three 2QL materials display an abrupt increase in $M(E)$ and $D(E)$ at the band edge. However, in this case the magnitudes of these discontinuities are not nearly as large as those of their 1QL counterparts. This is particularly evident with 2QL Bi$_2$Se$_3$ and Sb$_2$Te$_3$, due to their rings having small radii and thus containing fewer states. The starfish-shaped band of 2QL Bi$_2$Te$_3$ includes many states within a small energy range and results in an appreciable increase in $M(E)$ (although roughly half the value of 1QL Bi$_2$Te$_3$).

After their small but abrupt initial rise, the $M(E)$ and $D(E)$ distributions of 2QL Bi$_2$Se$_3$ and Sb$_2$Te$_3$ remain roughly constant until approximately $-0.1$~eV, at which point we observe sharp increases due to the ``starfish'' feature that contributes many states. With 2QL Bi$_2$Te$_3$, however, $M(E)$ and $D(E)$ remain mostly constant throughout the relevant energy range for transport. $V_{\lambda}(E)$ increases gradually from the band edge, then eventually drops with the presence of a local maxima ({\it e.g.} the starfish feature at roughly $-0.1$~eV in 2QL Bi$_2$Se$_3$ and Sb$_2$Te$_3$). As mentioned earlier, such features contribute a large number of low-velocity states that drag down $V_\lambda(E)$.

Next, we analyze the thermoelectric properties presented in Fig.~\ref{fig:2QL_props}(d)-(f). Note that we focus here on the results of the DOS scattering model, which is believed to be the most physical and accurate (a comparison of all three scattering models for the 2QL materials is provided in the Supplemental Information). The transport distributions for 2QL Bi$_2$Se$_3$ and Sb$_2$Te$_3$ initially rise, but suddenly decrease below $-0.1$~eV. This occurs as a result of the large DOS, and thus scattering, associated with the starfish feature. Similar behavior is observed with 2QL Bi$_2$Te$_3$, however the decrease is more gradual and begins below roughly $-0.15$~eV; this gives a higher $\Sigma(E)$ over a larger energy range compared to 2QL Bi$_2$Se$_3$ and Sb$_2$Te$_3$. These drops in $\Sigma(E)$, due to sudden increases in DOS, negatively affect the Seebeck coefficient and conductivity.

The power factor and $ZT$ of 2QL Bi$_2$Te$_3$ are higher than 2QL Bi$_2$Se$_3$ and Sb$_2$Te$_3$, because the increase in DOS from the low-energy ring is located far enough from the Fermi level that it has a minimal negative impact, unlike the starfish features. A comparison of the different scattering models (see the Supplemental Information) shows that the $ZT$ values are somewhat similar, but that the MFP and TAU models predict larger $PF$ peaks deeper in the valence band -- such peaks are suppressed with the DOS models due to the large number of states at lower energies. Finally, comparing the 2QL materials to their 1QL counterparts, the latter present overall better TE performance as a result of their ring-shaped dispersion which is mostly lost when going to 2QL.

\section{Conclusions}
\label{sec:summary}
In this study we employed first-principles modeling to calculate the thermoelectric properties of two-dimensional single and double quintuple-layer Bi$_2$Te$_3$, Bi$_2$Se$_3$ and Sb$_2$Te$_3$ using the MFP, TAU and DOS scattering models. The focus was to investigate how the different scattering approximations impact the TE characteristics of these materials, which possess unusual ring-shaped electron dispersions -- with an emphasis on the DOS scattering model.

The single QL materials display two ring-shaped valence band maxima between which is nestled one ring-shaped local minimum ({\it i.e.} the ``moat'' feature). The ring-shaped states at the band edge result in discontinuous increases in $M(E)$ and $D(E)$ that benefit TE performance -- as pointed out in previous studies \cite{Maassen2013,Wick2015,Zhou2015}. Interestingly, below the band edge, the moat produces an abrupt drop in $D(E)$ as those states are removed. This feature further enhances the power factor and $ZT$ with the DOS model, as a result of lower electron scattering compared to either MFP or TAU models. To confirm the role played by the two ring-shaped maxima and single ring-shaped minimum, we introduced an analytical octic band model that generally resembles the DFT dispersion. The octic model is found to reproduce the observed discontinuities in the DFT-computed $M(E)$, $V_\lambda(E)$ and $D(E)$.

Among the 1QL materials, Bi$_2$Te$_3$ showed the largest enhancement in TE performance with the DOS scattering model, as a result of the discontinuities brought about by the ring-shaped maxima and moat. Our findings indicate that the benefits of these unusually-shaped bands are sensitive to the relative alignment of the ring maxima and to the degree of ring anisotropy -- the latter smoothes out the desired abrupt features in the transport distribution. For these reasons, 1QL Bi$_2$Se$_3$ and 1QL Sb$_2$Te$_3$ do not show as much improvements in $PF$ and $ZT$ with the DOS model, compared to 1QL Bi$_2$Te$_3$.

With the 2QL materials, the electron dispersions are qualitatively different than their 1QL counterparts. 2QL Bi$_2$Te$_3$ shows a star-shaped dispersion near the band edge, with a ring feature near $-0.15$~eV. 2QL Bi$_2$Se$_3$ and 2QL Sb$_2$Te$_3$ both possess a small ring near the band edge, with starfish-shaped bands at roughly $-0.1$~eV. Due to the relatively large misalignement in energy of these features, lack of a distinctive minimum ({\it i.e.} moat), and the small radius of the rings at the band edge, these materials present significantly lower $PF$ and $ZT$ compared to the 1QL case. Double QL Bi$_2$Te$_3$ shows the best TE performance, among the 2QL materials, since the star-shaped dispersion provides a large number of states at the band edge, similar to a ring-shaped band.

Assuming that the DOS scattering model is the most physical and accurate, among the three models considered, our findings suggest that the 1QL form of Bi$_2$Te$_3$ and Bi$_2$Se$_3$, and to a lesser extent 2QL Bi$_2$Te$_3$, may provide better power factors than previous studies have suggested. We also show that a ring-shaped minimum below the band edge, which decreases $D(E)$ and thus scattering, can benefit TE performance. Further study of rigorous scattering in ring-shaped dispersions is needed to validate the accuracy of the DOS model for these materials. The results of this work help provide guidance for the search of other warped-band materials that may lead to exceptional TE characteristics.

\begin{acknowledgments}
This work was partially supported by DARPA MATRIX (Award No. HR0011-15-2-0037) and NSERC (Discovery Grant RGPIN-2016-04881). C. R. acknowledges support from an NSERC Canada Graduate Scholarship and Nova Scotia Graduate Scholarship.  
\end{acknowledgments}


\end{document}